\documentclass[journal]{IEEEtran}
\usepackage{tabularx}
\usepackage{verbatim}
\usepackage{arydshln}

\usepackage{cite}

\ifCLASSINFOpdf
  \usepackage[pdftex]{graphicx}
\else
  \usepackage[dvips]{graphicx}
\fi
\newcommand{\jn}[1]{\textcolor{magenta}{(JN: #1)}}
\usepackage{kotex}

\usepackage{booktabs}
\usepackage{multirow}
\usepackage{hhline}

\usepackage{amsmath}
\usepackage{amsfonts}

\usepackage{array}

\usepackage{comment}

\usepackage{stfloats}
\usepackage{url}

\usepackage[dvipsnames]{xcolor}

\newcommand{\tg}[1]{\textcolor{violet}{(Tg: #1)}}

\usepackage{hyperref}

\begin{document}
%

\title{Towards Efficient and Real-Time Piano Transcription Using Neural Autoregressive Models}





%
%
%

\author{Taegyun~Kwon,~\IEEEmembership{Member,~IEEE,}
        Dasaem~Jeong,~\IEEEmembership{Member,~IEEE,}
        and~Juhan~Nam,~\IEEEmembership{Member,~IEEE}
\thanks{T. Kwon and J. Nam are with the Graduate School of Culture Technology, Korea Advanced Institute of Science and Technology, Daejeon, South Korea, 34141, e-mail: juhan.nam@kaist.ac.kr.}
\thanks{D. Jeong are with Sogang University, Seoul, South Korea, 04107}
}

%
%

\markboth{Preprint Version (under review)}%
{Kwon \MakeLowercase{\textit{et al.}}: Efficient and Compact Autoregressive Model for Online Piano Transcription}
%



\maketitle

\begin{abstract}

In recent years, advancements in neural network designs and the availability of large-scale labeled datasets have led to significant improvements in the accuracy of piano transcription models. However, most previous work focused on high-performance offline transcription, neglecting deliberate consideration of model size. The goal of this work is to implement real-time inference for piano transcription while ensuring both high performance and lightweight. To this end, we propose novel architectures for convolutional recurrent neural networks, redesigning an existing autoregressive piano transcription model. First, we extend the acoustic module by adding a frequency-conditioned FiLM layer to the CNN module to adapt the convolutional filters on the frequency axis. Second, we improve note-state sequence modeling by using a pitchwise LSTM that focuses on note-state transitions within a note. In addition, we augment the autoregressive connection with an enhanced recursive context. Using these components, we propose two types of models; one for high performance and the other for high compactness. Through extensive experiments, we show that the proposed models are comparable to state-of-the-art models in terms of note accuracy on the MAESTRO dataset. We also investigate the effective model size and real-time inference latency by gradually streamlining the architecture. Finally, we conduct cross-data evaluation on unseen piano datasets and in-depth analysis to elucidate the effect of the proposed components in the view of note length and pitch range.        

\end{abstract}


%
\IEEEpeerreviewmaketitle

\section{Introduction}
%
%
%
%

\IEEEPARstart{A}{utomatic} music transcription (AMT) is a family of tasks that convert musical audio signals into a piano roll or musical notes \cite{benetos2018automatic}. Among others, the transcription of piano solo music has been the best-studied task because the note boundaries are well defined in time and it is relatively easy to obtain a large amount of training data with strongly labeled MIDI by synthesizing virtual piano sounds from MIDI data \cite{emiya2010maps} or by capturing MIDI data from piano performance actions on computer-controlled pianos such as the Disklavier \cite{hawthorne2018enabling}. \emph{Onsets and Frames} is the milestone model that achieved a very high note accuracy (around 95\% in note-level F-score) by harnessing the power of deep neural networks and the abundance of strongly labeled training data \cite{hawthorne2017onsets}.

\begin{figure}[!tp]
\centering
 \includegraphics[width=\columnwidth]{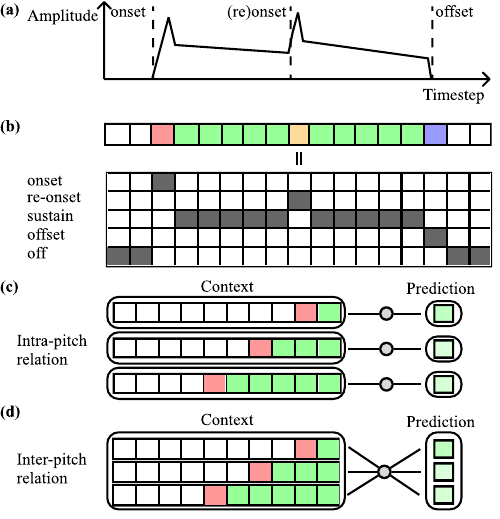}
 \caption{Frame-level note states and their relations across different pitches. (a) A simplified amplitude envelope of two successive notes while the sustain pedal is pressed. (b) Multi-state representation of notes. In our model, each note is in one of the five states. (c) \textit{Intra-pitch} relation considers note states within a single pitch only, (d) \textit{Inter-pitch} relation considers note states across all pitches.  We constrained our model to have only intra-pitch relations by removing the inter-pitch link in LSTM.
 }
 \label{fig:notes}
\end{figure}

A common architecture of neural network models in the \emph{Onsets and Frames} and the following work is convolutional recurrent neural networks (CRNN) where the convolutional modules capture local acoustic features from the mel spectrogram and the recurrent modules use the local features with temporal connections to detect note onsets or frame-level on/off states (frames) \cite{kimadversarial, hawthorne2017onsets, kelz2019adsr, kong2021high} in a broad context. While they have shown superior performance, we observe two main limitations. 
First, they use the CRNN model to detect note states through binary classification. This approach requires multiple CRNN branches to independently predict note onset, offset, and frame (on/off state), resulting in potential redundancy within the model. Second, the bidirectional connections in the recurrent modules allow the model to capture a broader temporal context but hinder real-time inference, limiting the practical applications of piano music transcription \cite{carabias2013constrained, pesek2017robust, vaca2019real, onpotential}. While removing the backward connection provides a simple solution, it significantly degrades performance (about 2-3\% in F-score) \cite{hawthorne2017onsets}. To address these challenges, we previously proposed an autoregressive multi-state note model \cite{kwon2020polyphonic}. The model has a softmax output that predicts multiple note states (onset, offset, sustain, re-onset, off) from a single CRNN branch as shown in Figure \ref{fig:notes} (a) and (b)) rather than using separate binary outputs from multiple branches, thus potentially reducing model size. In addition, the model resolved the real-time inference problem by incorporating an autoregressive connection into a unidirectional recurrent module, allowing the model to learn a broader context during real-time note inference \cite{Jeong2020}. In this paper, we redesign the autoregressive multi-state note model for lightweight and real-time inference while improving the performance further.

The first refinement pertains to CNN blocks. In recent years, a standard prototype for this purpose has been a simple VGG-style CNN configuration with 3x3 filters on mel spectrograms, as used in various studies \cite{onpotential, kimadversarial, kelz2019adsr, wang2018polyphonic, hawthorne2018enabling, kwon2020polyphonic}. However, the conventional 2D convolutional layers assume the filters are translation invariant on both the time and log-scaled frequency axis. In reality, time-frequency patterns exhibit variations over different frequency ranges; for example, high-pitched notes exhibit faster amplitude decay than low-pitched notes. While increasing the number of filters can help to accommodate this variation, it also leads to a larger model size. To effectively capture frequency dependencies, we introduce frequency-conditioned Feature-wise Linear Modulation (FiLM) layers to the feature maps of the CNN blocks. Considering that local acoustic feature properties change continuously with pitch height, we integrate FiLM layers to modulate the feature maps along the frequency axis. 

The second refinement focuses on RNN blocks. We use an autoregressive LSTM to learn the temporal relations of frame-level note states. Since the model takes input and output over the entire 88 pitches in an autoregressive manner, it captures not only the temporal relations of frame-level note states (e.g., onset, sustain, and offset) within a single note, but also the harmonic or sequential relations between different notes. These relations are referred to as \emph{intra-pitch} and \emph{inter-pitch}, respectively, as illustrated in Figure \ref{fig:notes} (c) and (d). Learning the \emph{inter-note} relation is similar to language modeling in the domain of symbolic music, which is the core concept in symbolic music modeling \cite{boulanger2012modeling}. Leveraging such musical knowledge can potentially improve music transcription in a more musically meaningful manner. However, Ycart et al. have reported that integrating language models may not be effective because it requires learning high complexity and dealing with long-term dependencies when the sequence is given at the frame level \cite{Ycart2019BLENDINGAA}. Our own research has also shown that note decoding based on the musical language model is computationally expensive and does not outperform a simple rule-based decoding approach \cite{kwon2020polyphonic}. Based on this observation, the autoregressive model prioritizes \emph{intra-pitch} relations while neglecting \emph{inter-pitch} relations. To achieve this, we incorporate 88 pitch-wise LSTMs, each operating independently on every key of the piano. Furthermore, to significantly reduce the number of parameters, we ensure that these 88 pitch-wise LSTMs share the same parameters. While this approach requires a large RNN capacity to account for the variation across the entire pitch range, we assume that the fundamental role of the RNN, which involves learning the temporal relations at the frame level, remains relatively consistent across different pitches. In addition, the frequency-conditioned feature maps produced by the FiLM layers in the CNN module could compensate for any potential lack of pitch dependence in the model.

We propose two types of models using the new CNN and RNN components; one for high performance and the other for high compactness. We validate the two models through comprehensive experiments including an ablation study to show the effectiveness of individual components, different model configurations to investigate the model size and latency, and evaluation on unseen piano music datasets. We also analyze the prediction errors of the models in the view of note length and pitch range to better understand the effect of the CNN and RNN components. We present piano transcription examples and video demos for real-time inferences at the companion website \url{https://taegyunkwon.github.io/PARpiano/}. 

\section{Related Work}
AMT is analogous to automatic speech recognition (ASR) in that it predicts a symbolic sequence from acoustic signals. Since ASR systems typically consist of an acoustic model and a language model, AMT researchers have borrowed the concept and have designed the two models separately or jointly. We follow the same approach, but instead of having an explicit note-level language model, we use a note-state sequence model, where the sequence model plays a limited role, learning the transition of note states at the frame level.  
We review previous work in this perspective and contrast them to our proposed model. 

\subsection{Acoustic Model}
The acoustic model in AMT typically takes a spectrogram as input and extracts local time-frequency patterns to predict frame-level note activations. Early work used non-negative matrix factorization (NMF) \cite{vincent2009adaptive} or support vector machine (SVM)\cite{nam2011, poliner2006discriminative}. Kelz et al.\cite{onpotential} showed that high performance can be achieved by a simple CNN model consisting of convolutional layers with 3x3 filters, max-pool layers, and fully connected layers. Since then, this CNN configuration has been used as a standard prototype \cite{kong2021high, kimadversarial, hawthorne2018enabling, kwon2020polyphonic, yan2021skipping}. Other variations include the use of filters with different sizes  \cite{kelz2019adsr, Two-Stage} and U-net \cite{wu2020multi}, or the use of CQT instead of mel spectrogram \cite{Two-Stage, taenzer2021informing}. The CNN architectures use 2D convolutional layers in common. They assume that the filters are translation invariant not only along the time axis, but also the log-scaled frequency axis. However, in musical signals, low and high pitch ranges have different patterns of acoustic characteristics. In the case of piano, the timbre is different in the low and high registers. This is usually ignored in most  CNN designs. In this study, we address this issue by applying the frequency-conditioned FiLM layers to the feature maps in the CNN module. This modulates the time-frequency feature maps along the frequency axis, thus adapting the extracted features to the pitch-dependent characteristics. This also compensates for the limited capacity of the single RNN module, which shares parameters across the entire pitch range in the note-state sequence model. 

\begin{figure}[!t]
\centering
 \includegraphics[width=\columnwidth]{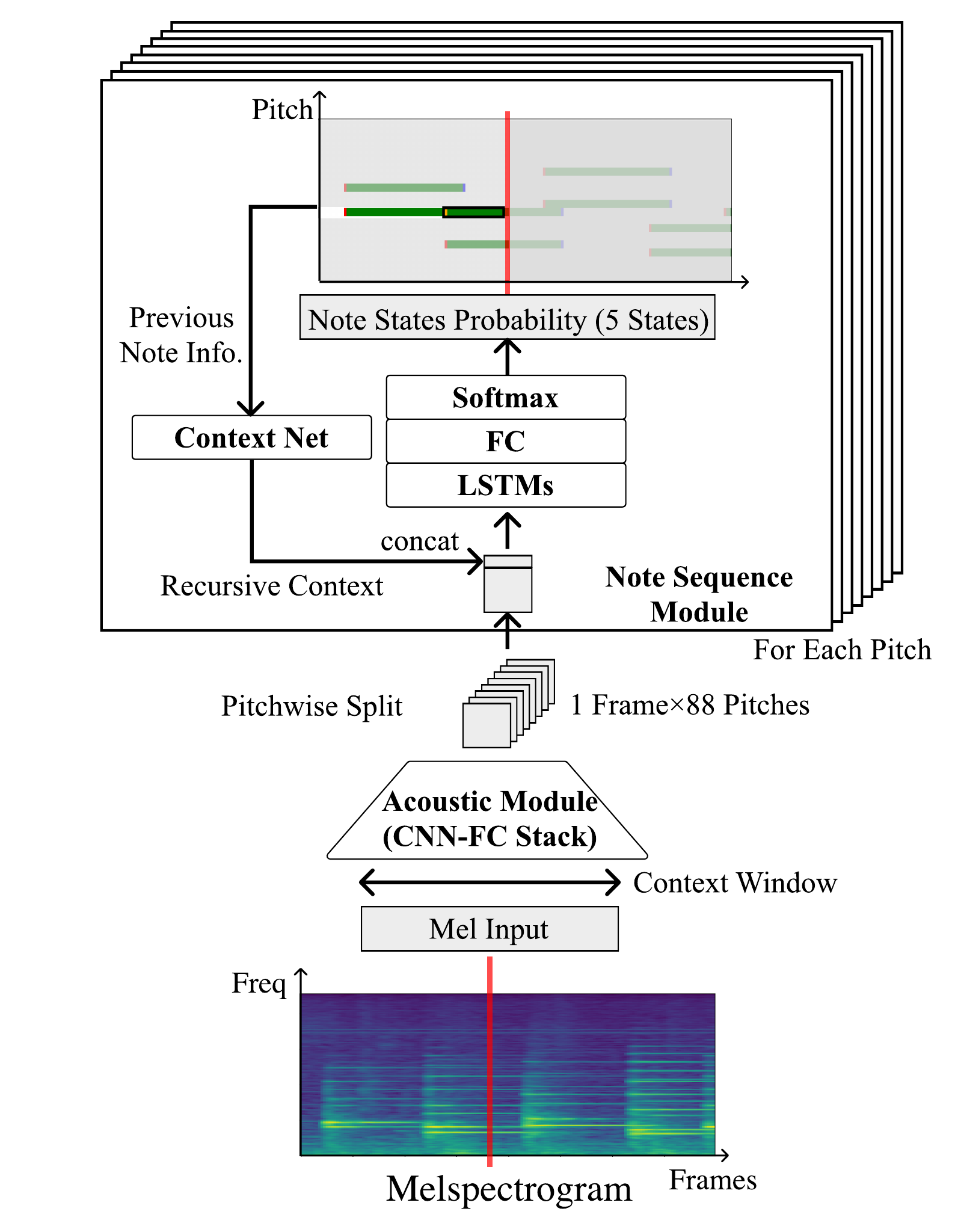}
 \caption{\textbf{Overall system diagram.} A step prediction of the system is illustrated with the context window required for single step. The system takes $6$ more frames in forward direction, which corresponds to 320 ms. The output of the acoustic model are split into 88 segments, corresponding to 88 pitches. Each segment is processed separately and then combined together. For simplicity, velocity modules are omitted.} 
 \label{fig:overall}
\end{figure}

\subsection{Musical Language Model}
\label{subsec:language_models}
Unlike ASR, the output in AMT is time-aligned to the input audio and is usually represented at the frame level in the form of piano rolls. The language model should therefore learn the note activation patterns in a very high-dimensional space (e.g., 88$\times$the number of frames in a piano roll). 
Previous work has largely tackled this problem using one of three approaches. The first is to learn the correlation of note activations across the entire high-dimensional space. Early work used restricted Boltzmann machines (RBM), RNN and neural autoregressive distribution estimator (NADE) \cite{boulanger2013transduction, wang2018polyphonic}. More recent work has used adversarial discriminator \cite{kimadversarial}, autoregressive RNN \cite{Ycart2019BLENDINGAA}, and autoregressive transformer \cite{seq2seq}. In general, these models are computationally expensive and require a highly complex decoding scheme such as beam search. Some of them mitigate the complexity by converting the frame-level input to a note-level input or an event-based input \cite{Ycart2019BLENDINGAA,seq2seq}. 
The second approach is to learn the correlation of note activations pitch by pitch. In other words, the model treats the note states at each pitch independently. This approach prevents the model from learning the inter-note relations and therefore the music language model focuses on the temporal evolution of each note. This approach has been used as a post-processing (or smoothing) method on the frame-level note activations, mainly based on the hidden Markov model (HMM) \cite{poliner2006discriminative, nam2011, kelz2019adsr}. The last approach uses only the acoustic model without any explicit language model. The \emph{Onsets and Frames} and the subsequent work showed that a simple rule-based inference on the output of the acoustic model is sufficient if the acoustic model is designed to detect different note states well \cite{hawthorne2017onsets,kong2021high}. 

In our previous work, we set the CRNN model to cover the entire high-dimensional pitch activation space in an autoregressive manner (the first approach) \cite{kwon2020polyphonic}. As a result, the note decoding based on beam search was very complex and did not even perform better than a simple rule-based decoding. In this work, we redesign the RNN module to take the high-dimensional pitch activations pitch by pitch, thus discarding the interactions between different pitches (the second approach). In addition, we set the 88 pitch-wise RNNs to share the same parameters. This significantly reduces the number of parameters, making the model lightweight.

\section{Method}
\label{sec:method}


This section introduces the proposed model and describes the details of the model design.  

\subsection{Model Overview}
An overview of the proposed model is shown in Figure \ref{fig:overall}. It is based on the autoregressive multi-state note model from our previous study \cite{kwon2020polyphonic}. The two main components of the base model are the recursive connection from the note states in the previous time step to the output of the acoustic module in the unidirectional RNN module, and the multi-state softmax output (i.e. onset, sustain, re-onset, offset, and off), which allows the model to have a single CRNN branch. 
The base model is modified in both the CNN and RNN modules. First, frequency-conditioned FiLM layers are added to the CNN module to increase the expressive power of the acoustic module in the frequency dimension. 
The feature maps from the CNN module are divided into 88 groups across the channel. In the RNN module, each of the feature map groups is processed separately by the pitch-wise RNNs.  
The pitch-wise RNNs are composed of 88 LSTMs and they share the parameters. With these new components, we present two specific models. One is the ``Pitch-wise AutoRegressive (PAR)'' model, which is relatively large in model size but has high performance. The other is the ``PAR$_{Compact}$'' model, which is much smaller in model size but maintains decent performance. Both models require 6 more frames in the forward direction. This causes a latency of 320 ms in the real-time inference. This set was close to the tolerance (300 ms) in the evaluation of score following systems \cite{cont2007evaluation}. 

\begin{figure*}[!htp]
 \includegraphics[width=\textwidth]{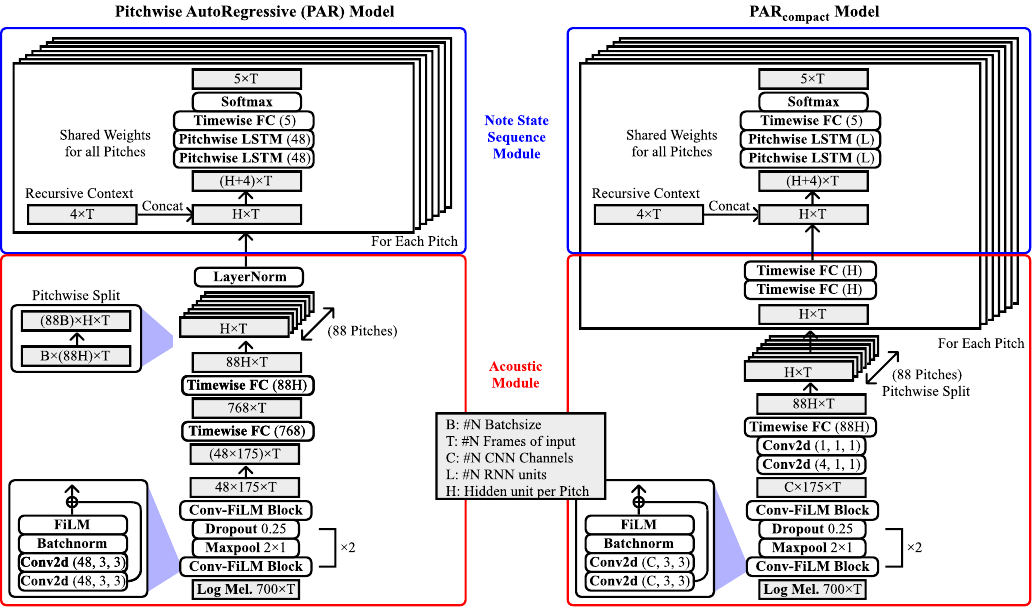}
 \caption{\textbf{Model Diagram.} The gray boxes indicate feature maps, and the rounded white boxes indicate operations. The numbers in the Conv2d block refer to ($channels, filter\ width, filter\ height$) respectively. The Timewise FC refers to the application of a fully connected layer at each time step.
 The PAR$_{Compact}$ model differs from the PAR model only in the middle part, as it reduces the number of channels by applying 1x1 convolution, and splits the feature maps into 88 pitches earlier to avoid having fully connected layers with a large number of parameters. 
}
 \label{fig:model}
\end{figure*}

\subsection{Acoustic Module}
\label{subsec:acoustic}


The model structure is shown in Figure \ref{fig:model}. The acoustic module was set up to take mel spectrogram with 700 mel bins because our preliminary experiments show that the large size gave better performance than the commonly used 229 bins \cite{kong2021high, hawthorne2017onsets}.
The mel spectrogram is processed by three Conv-FiLM blocks and two max-pooling layers. The Conv-FiLM block is composed of 3x3 convolutions and a FiLM layer. The max-pooling layer is placed between the Conv-FiLM blocks and reduces the frequency dimension by 2. All channels of the resulting feature maps are combined (indicated as 48x175 ) and mapped to 768 neurons per frame through a time-wise fully connected (FC) layer.
The output is then mapped by another time-wise FC layer with a multiple of 88 nodes, and the resulting feature maps are split into 88 pitch segments of the same size, which are then passed to the note state sequence module.

\subsubsection{Frequency-conditioned FiLM}
\label{subsec:film}
Typical CNN modules used for a transcription model consist of several convolutional layers with small 2D kernels (3x3) and a fully connected layer. One drawback of such a design is that all frequency ranges are processed with the same filters. This setup is good for learning common time-frequency patterns over the entire frequency range, but ignores differences between low and high frequencies, such as different decay times of overtones. We compensate for this limitation by adding learnable linear modulation via the FiLM layer. 


Given a feature map $X$ of a convolution layer, where $i$ is the channel index, the FiLM layer provides the feature map with a channel-wise transformation on a condition $c$ as follows:
\begin{equation}
    FiLM(X_{i}|c) = \gamma_{i,c} X_{i} + \beta_{i,c}.
\end{equation} 
where $\gamma$ and $\beta$ are parameterized by a small trainable neural network. In the original paper \cite{perez2018film}, the FiLM layer was used to transform the feature map of the visual module according to the given question in the visual question-answer task. 
We modified the FiLM layer to transform the feature map based on the relative frequency height of the feature map of the Conv2d layer, which corresponds to the continuous pitch change from low to high.   
Specifically, we set the relative frequency height as $k/F$ where $F$ is the size of the feature map in height and $k$ is the height index mapped to the continuous pitch change. 
Both $\gamma$ and $\beta$ take $k/F$ as the pitch condition and modulate the feature maps via a small neural network ($f,g$) (the channel index is omitted) as shown below: 
\begin{equation} 
FiLM(X_{k}) = \gamma_k X_k + \beta_k = f(\frac{k}{F}) \cdot X_k + g(\frac{k}{F})
\end{equation}
We used four layers of fully connected networks with the ReLU activation for $\gamma$ and $\beta$. The size of the last layer was equal to the number of the channels in the feature map, so that each index corresponds to each channel.

\subsection{Note-State Sequence Module}
\subsubsection{Pitch-wise LSTM}
\label{subsec:pitchwise}

The note-state sequence module takes the 88 segments from the acoustic module output and processes them independently with pitch-wise LSTMs. The pitch-wise LSTMs focuses on the temporal transition within a pitched-note (\textit{intra-pitch}) rather than learning relationships between multiple notes (\textit{inter-pitch}). The same pitch-wise LSTMs are used across the entire 88 segments, assuming that the temporal evolution within a note is independent of pitch. This parameter sharing significantly reduces the model size. The output of pitch-wise LSTMs are projected into a five-dimensional vector using the softmax layer where each dimension corresponds to the probability of the note state (onset, re-onset, sustain, offset, and off) at a specific pitch and time step.

\subsubsection{Enhanced Recursive Context}
\label{subsec:ARconnection}

\begin{figure}[!tbp]
 \includegraphics[width=\columnwidth]{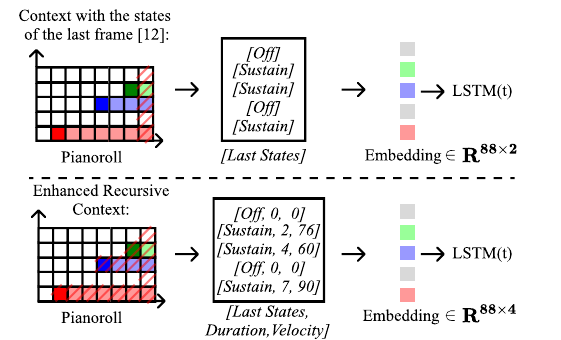}
 \caption{\textbf{Enhanced Recursive Context.} The left piano roll depicts notes states at the specific step $t-1$ and the highlighted areas indicate the context being considered. In \cite{kwon2020polyphonic}, only the last states were considered as a context and converted to a 2-dimensional embedding per pitch. In the Enhanced Recursive Context setting, we additionally consider note velocity and note duration. Also, the contexts are encoded into a 4-dimensional embedding. }
 \label{fig:contextnet}
\end{figure}

Our previous auto-regressive multi-state note model used note states at the last frame for the recursive connection in the LSTMs \cite{kwon2020polyphonic}. This single-frame context does not contain information about how long the activated notes have been sustained, which often resulted in overly long notes being predicted when the model missed the note offset state. 
To overcome this issue, we augment the recursive context so that the pitch-wise LSTMs catch up the states of the playing notes effectively. Specifically, we add the duration and velocity of the playing notes to the note states, as illustrated in Figure \ref{fig:contextnet}. The duration is counted as the number of lasting frames from the onset (starting from 1). The velocity is the MIDI note velocity of the playing notes. 
If a note is off, velocity and duration are simply set to 0. The values are normalized to the maximum value (for duration, we trimmed it to 5 secs) and embedded with two layers of small fully connected layer with 16 units, and projected into 4 dimension with a linear layer as a context vector. This is concatenated with the output of the acoustic module for each pitch as the input of the pitch-wise LSTMs. 


We expect that the duration and velocity information in the recursive context provides a more complete view of the latest notes helps the model to compare it to current acoustic features. For example, a note with a long duration and low velocity indicates that the sound intensity of the last note is weak, so the LSTM may become sensitive to detecting offset. Note that we used a teacher-forcing method to train the model. In the inference phase, the recursive context is calculated from the model prediction.  

\subsection{Compact Model}
We observed that the fully connected network on top of the Conv-FiLM blocks in the acoustic module took up most of the parameters in the PAR model. This issue is found in VGG or other CNN models that have large feature maps beneath the fully connected layers as well \cite{simonyan2015very}. To address the issue, we additionally propose a compact version of the PAR model. As shown in Figure \ref{fig:model}, the compact model reduces the number of channels of the output of Conv-FiLM blocks to one with two $1\times 1$ convolution layers, which was inspired by the inception module \cite{szegedy2015going}. Then, the feature map is split to 88 pitch segments. Lastly, the timewise FC layers above operate on each pitch feature separately, sharing the parameters. These modifications significantly reduce the model size. In the experiment section, we will investigate the size limit of the compact model by changing the number of channels in the Conv-FiLM blocks, the number of hidden units per pitch in the FC layers, and the number of LSTM units.

\vspace{1cm}
\section{Experiments}

\label{sec:experiments}
\subsection{Datasets}
\label{subsec:datasets}

{\color{blue}

}
We used five piano music datasets where audio and MIDI are paired. They are all different in terms of piano quality and recording environments. We used the MAESTRO dataset\cite{hawthorne2018enabling} as the primary one, as it has been used as a benchmark in recent years \cite{kong2021high, hawthorne2018enabling, seq2seq, kimadversarial,yan2021skipping}. We trained the model only with the training split of the MAESTRO dataset and evaluated the model with the rest. We provide a concise overview of the five datasets below.


\subsubsection{MAESTRO \cite{hawthorne2018enabling}}
This is the largest public dataset for piano music transcription. It contains piano performances played in \textit{International Piano-e-Competition}\footnote{www.piano-e-competition.com}. The MIDI files were recorded through \textit{Yamaha Disklavier} pianos which can capture MIDI messages via built-in sensors. The dataset has three versions. We used the first (V1) and third (V3) versions for the performance comparison. The V3 is an updated version of the V2, removing six recordings with string quartet accompaniment that were mistakenly included.


\subsubsection{MAPS \cite{emiya2009multipitch}}
This contains 7 sets of synthesized piano and 2 sets of acoustic piano (\textit{Yamaha Disklavier Mark III}). We used only the acoustic piano sets which include a total of 60 pieces in the `MUS' subset. 


\subsubsection{Vienna 4x22 Corpus \cite{vienna4x22}}
This contains recordings of 4 short classical pieces played by 22 pianists on the Bösendorfer SE290 grand piano. Since the provided MIDI was not aligned with audio, we manually aligned it.  

\subsubsection{Saarland music dataset (SMD)}
We used the piano subset of this dataset \cite{MuellerKBA11_SMD_ISMIR-lateBreaking}. It contains 50 performances played on a Yamaha Disklavier model by students of piano classes of different levels. 

\subsubsection{In-house Dataset}
We collected our own recordings of piano performances on \textit{Yamaha Disklavier C7 Enspire}. The recorded pieces consist of brief excerpts (approximately 30 s in length) extracted from classical piano compositions. These performances were performed by highly proficient undergraduate students majoring in piano performance. This dataset contains 461 recordings in total.

Note that we elongated the note offsets when the sustain pedal is pressed, using the control value of 64 as a threshold \cite{kwon2020polyphonic, hawthorne2017onsets}. However, on the SMD dataset, we observed the effective threshold was in a lower range. After a grid search in the evaluation phase, we selected a threshold value of 21 instead of 64. 


\subsection{{Evaluation Metrics}}
We used standard note-level evaluation metrics including precision (P), recall (R), and F1-score (F1) and calculated them using the \textit{mir\_eval} library \cite{mir_eval}. 
Following the previous work, note onsets were counted as correct if they are within $\pm$50 ms of the ground truth and note offsets were as correct if they are within  $\pm$50 ms of the ground truth or $\pm$20\% of the note duration For note velocity, we used $\pm$10\% of the ground truth velocity as a tolerance window.  
In addition to the standard metrics, we calculated ``note duration accuracy'' to measure the accuracy of note offset given the correct note onset. It was obtained by taking the ratio of the recall of note offset to the recall of note onset.  
For the previous studies where we cannot directly calculate note duration accuracy, we approximated the value by dividing the overall offset recall by the onset recall (which introduces about $\pm 0.1\%p$ approximation error in our own results).

\label{subsec:overall}
\begin{table*}[!htp]
\caption{\textbf{Overall Results} The parameters number of Multi-State Model \cite{kwon2020polyphonic} was counted from the model without note velocity prediction. 
}
\label{tab:overall}

\resizebox{\textwidth}{!}{
\begin{tabular}{@{}llcrcccccccccc@{}}
\toprule
 &                    & \multicolumn{1}{c}{Online} & \multicolumn{1}{c}{Params} & \multicolumn{3}{c}{Note} & \multicolumn{3}{c}{\begin{tabular}[c]{@{}c@{}}Note w/ offsets\end{tabular}} & \multicolumn{3}{c}{\begin{tabular}[c]{@{}c@{}}Note w/ Offset \\ $\And$ Velocity\end{tabular}} & \begin{tabular}[c]{@{}c@{}}Note Duration\\ Accuracy\end{tabular}\\ \midrule
 &            &        & \multicolumn{1}{c}{}          & R (\%)        & P (\%)         & F1 (\%)         & R (\%)          & P (\%)                      & F1 (\%)                         & R (\%)                           & P (\%)                          & F1 (\%)             & \%                  \\ \midrule
\multicolumn{1}{c|}{\multirow{5}{*}{\rotatebox[origin=c]{90}{\parbox[c]{1.65cm}{\centering MAESTRO V1}}}} 
&    Adversarial\cite{kimadversarial} &  &  26.9M                         & 93.2      & 98.1      & 95.6     & 79.3                        & 83.5                          & 81.3                       & 78.2                            & 82.3                              & 80.2   & 85.1$^{\#}$                         \\
\multicolumn{1}{l|}{}                            & Onsets and Frames \cite{hawthorne2018enabling}    &            & 24.7M                        & 92.6     & 98.3     & 95.3    & 78.2                       & 83.0                         & 80.5                       & 75.4                           & 79.9                             & 77.5    & 84.5$^{\#}$                      \\
\multicolumn{1}{l|}{}                            & Seq2Seq\cite{seq2seq} &  & 54M                         &  \textemdash  &   \textemdash & 96.0   &               \textemdash     &         \textemdash         & 83.5                     &            \textemdash    &      \textemdash    & 82.2         & \textemdash    \\  
\multicolumn{1}{l|}{}                            & Multi-State Model\cite{kwon2020polyphonic}   &\checkmark           & 14.6M$*$                        & 91.2     & \textbf{98.7}     & 94.7    & 76.5                       & 82.6                         & 79.4                       &      \textemdash                  &         \textemdash     &    \textemdash  & 83.9                        \\
\multicolumn{1}{l|}{}                            & Proposed (PAR)   &\checkmark  &   19.7M  &  \textbf{95.6} & 98.4 & \textbf{96.9} & \textbf{86.5} & \textbf{89.0} & \textbf{87.7} & \textbf{85.3} & \textbf{87.7} & \textbf{86.5} &  \textbf{90.6}             \\
\multicolumn{1}{l|}{}                            & Proposed (PAR$_{compact}$)   &\checkmark  & 2.7M  &  94.9 & 97.2 & 96.0 & 83.7 & 85.7 & 84.7 & 70.5 & 81.5 &  80.6  & 88.3                  \\
 \midrule
\multicolumn{1}{c|}{\multirow{7}{*}{\rotatebox[origin=c]{90}{\parbox[c]{1.3cm}{\centering MAESTRO V3}}}} & Onsets and Frames [reproduced]      &           & 24.7M                        & 89.4     & \textbf{99.5}     & 94.1    & 75.9                       & 84.2                         & 79.8                       & 72.5                           & 80.3                             & 76.1       & 84.9                   \\ 
\multicolumn{1}{l|}{}                            & Onsets and Frames [reproduced]  & \checkmark   & 20.9M                        & 88.8     & 99.4     & 93.7    & 73.9                       & 82.5                         & 77.8                       & 70.6                           & 78.8                             & 74.4           & 83.2               \\

\multicolumn{1}{l|}{} & Regressing Onsets.\cite{kong2021high}    &      & 20.2M   & \textbf{95.6}     & 98.1     & 96.8    & 83.7                       & 85.8                         & 84.7                      & 82.2                            & 84.3                              & 83.3 & 87.2\\ 

\multicolumn{1}{l|}{} & Seq2Seq\cite{seq2seq}  &     & 54M                          &    \textemdash    &     \textemdash    & 96.0    &     \textemdash         &         \textemdash           & 83.9                      &       \textemdash        &         \textemdash       & 82.8        & \textemdash                   \\
\multicolumn{1}{l|}{} & Neural CRF\cite{yan2021skipping}        &      &              9M   &  94.0     & 98.7      & 96.1    &  86.5                     &  \textbf{90.8}                        & \textbf{88.4}                      & \textbf{85.5}                     &    \textbf{89.7}                          & \textbf{87.4}       & \textbf{92.0}$^{\#}$                    \\
\multicolumn{1}{l|}{}                            & Proposed (PAR) & \checkmark & 19.7M                        & \textbf{95.6} & 98.5 & \textbf{97.0} & \textbf{86.7} & 89.2 & 87.9 & \textbf{85.5} & 88.1 & 86.8     & 90.7                         \\ 
\multicolumn{1}{l|}{}                            & Proposed (PAR$_{compact}$) & \checkmark  & 2.7M                         & 94.3 & 96.9 & 95.5 & 83.9 & 86.1 & 85.0 & 83.9 & 86.1 & 85.0                               & 89.0                   \\
\bottomrule

\end{tabular}
}
\end{table*}

\subsection{Training Details}
\label{sec:hparam}

We used audio with a sample rate of 16 kHz and computed a log mel spectrogram based on a short-time Fourier transform with a window size of 4096 and a hop size of 512, covering frequencies from 27.5 Hz to 8372 Hz, corresponding to the lowest and highest pitches of a piano. The number of mel bins was set to 700. The overall configuration of the proposed model is described in Figure \ref{fig:model}. The PAR model has 48 channels for all convolution layers, 768 nodes for the fully connected layers, and 48 units for the pitch-wise LSTM layer. For the modulation functions $\alpha$ and $\beta$ in the frequency-conditioned FiLM, we used three FC layers with 16 hidden units. The window size ($W$) of the multistep FC layer was set to 5. For the compact model, we adjusted the size of the convolution, FC, and LSTM layer to progressively reduce the model size. We tested a total of 5 models that cover a wide range of model sizes. The number of parameters, unit size, and performance are compared in Table \ref{tab:PCmodel}. 


For estimating note velocity, we used another model with the same specifications as the note model, as done in the onsets and frames model \cite{hawthorne2017onsets}. The LSTM of the velocity model takes both the output of the acoustic module from the note model and the velocity model. To prevent interference between the two models, the gradient was disconnected between the velocity model and the note model.


All models were trained with the focal loss\cite{lin2017focal} with $\alpha=1.0, \gamma=2.0$ and the Adabelief optimizer \cite{zhuang2020adabelief} with an initial learning rate of 1e-3, and a batch size of 12. For note velocity, we linearly mapped the velocity range [0, 127] to [0,1] and applied the L2 loss only on the note onset frame of the ground truth. We randomly selected 10 s during training but we evaluated the entire piece at once during inference.
We selected a model with the highest F1 validation score in note offset for up to 250k iterations.
For the implementation, we used \textit{Pytorch 1.11} and trained the models on two NVIDIA GeForce RTX 2080 TI GPUs.

\section{Results}
\label{sec:result}

\subsection{Performance Comparison}
Table \ref{tab:overall} shows an overall comparison between the proposed model and recent state-of-the-art models evaluated on two versions of the MAESTRO dataset. Regarding the Onsets and Frames model, we reimplemented it to run both offline and online and evaluated it on MAESTRO V3. The online model was implemented by modifying the CRNN model with unidirectional LSTMs\footnote{The hyper-parameters were set in the model in \cite{hawthorne2018enabling}}. For the Regressing Onsets\cite{kong2021high} model, we re-evaluate it pre-trained model\footnote{We accessed the pre-trained model at \url{www.github.com/bytedance/piano\_transcription} on Dec 14, 2021} on MAESTRO V3 as the original paper reported the result evaluated on MAESTRO V2\footnote{We observed that the results on MAESTRO V3 in Table \ref{tab:overall} are higher than those evaluated on MAESTRO V2 in the original paper.}  
From the table, we can see that the proposed PAR model achieves higher or comparable accuracy in note and note with offset. This indicates that the state-of-the-art level of note onset and offset accuracy can be achieved without a backward context in the RNNs and learning inter-pitch relations. The PAR$_{compact}$ model also achieves comparable accuracy to other state-of-the-art models, demonstrating the effectiveness of the streamlined model.

Compared to our previous multi-state model\cite{kwon2020polyphonic}, the PAR model gained 2.2\%p and 8.3\%p on note onset and offset F1 scores, respectively. In addition, the number of notes with excessively long offset prediction where the predicted duration is more than 3sec longer than the actual duration decreased by about 70\% (from 1.08\% to 0.34\%). As a result, the PAR model shows improvement in note offset accuracy, which is confirmed by the note duration accuracy, although it is slightly lower than the note F1 score with offset in the Neural CRF model. In Section \ref{subsec:abalation}, we will show that this is the result of several contributing factors. 


\begin{table*}[ht!]
\caption{\textbf{Ablation Study - PAR Model.} The notation W/o indicates that the component was removed or replaced.}
\label{tab:PAR_ablation}
\resizebox{\textwidth}{!}{
\begin{tabular}{l|cccccccccc}
\toprule                                                                                                     & \multicolumn{3}{c}{Note} & \multicolumn{3}{c}{\begin{tabular}[c]{@{}c@{}}Note w/ offsets \end{tabular}} & \multicolumn{3}{c}{\begin{tabular}[c]{@{}c@{}}Note w/ offsets \\ $\And$ Velocities\end{tabular}}&{\begin{tabular}[c]{@{}c@{}}Note Duration\\ Accuracy\end{tabular}} \\ \hline
Model  & R (\%)    & P (\%)    & F1 (\%)       & R (\%)         & P (\%)        & F1 (\%)           & R (\%)              & P (\%)              & F1 (\%)  &  (\%)                \\ \hline
(a) W/o Pitchwise LSTM& 92.57 & 98.85 & 95.55 & 80.56 & 85.95 & 83.12 & 79.51 & 84.79 & 82.02  & 90.42    \\
(b) W/o Enhanced Context&  95.37 & \textbf{98.87} & \textbf{97.06} & 85.95 & 89.08 & 87.47 & 84.80 & 87.86 & 86.28   & 90.02 \\
(c) W/o Freq.-conditioned FiLM& 94.81 & 98.82 & 96.74 & 85.36 & 88.92 & 87.08 & 84.24 & 87.73 & 85.92   & 90.08        \\
(d) W/o $\{$Pitchwise, Enh. Context, FiLM$\}$&90.62 & 98.79 & 94.46 & 76.53 & 83.33 & 79.72 & 75.50 & 82.17 & 78.64  & 84.49          \\
(e) Non-Autoregressive &  93.17 & 97.98 & 95.47 & 82.32 & 86.46 & 84.30 & 81.04 & 85.09 & 82.97  & 88.41  \\
(f) PAR (All Components) & \textbf{95.60} & 98.49 & 97.01 & \textbf{86.63} & \textbf{89.21} & \textbf{87.88} & \textbf{85.52} & \textbf{88.05} & \textbf{86.75} & \textbf{90.67}    \\       
\hline
\end{tabular}
}
\end{table*}

\subsection{Ablation Study}
\label{subsec:abalation}
We conducted an ablation study to investigate the influence of each component in the PAR model (Table \ref{tab:PAR_ablation}). We examined the model by modifying the following components of the PAR model one by one: (a) replacing the pitch-wise LSTM layers by a global LSTM layer with 256 nodes covering the entire pitch range, (b) replacing the enhanced recursive context by a context with the states of the last frame only,  (c) removing the FiLM layers, (d) replacing or removing the three components (pitch-wise LSTM, enhanced recursive context, frequency-conditioned FiLM layer) together, (e) removing the autoregressive connection.


The ablation shows that the pitch-wise LSTM is a critical factor in improving performance on all accuracy metrics. Removing the enhanced context or frequency-conditioned FiLM alone does not affect the results much, as shown in (b) and (c), but when they are removed together with the pitch-wise LSTM, the performance drop is highly significant (d), greater than the sum of the individual performance drops from (a), (b), and (c). In particular, a significant difference in note duration accuracy indicates that the improvement of the PAR model in note offset metrics is not only due to the improvement in onset but also to independent improvements in note duration accuracy (offset). Finally, the autoregressive connection is as important as the pitch-wise LSTM in terms of accuracy scores, as shown in (e). 

\begin{figure}[!tp]
 \includegraphics[width=\columnwidth]{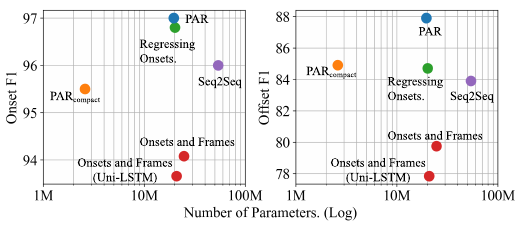}
 \caption{\textbf{Model Parameters vs. Note F1 Score.} }
 \label{fig:model_param}
\end{figure}
 
\begin{table}[!t]
\setlength{\tabcolsep}{3pt}
\caption{\textbf{Compact Model Experiment}}
\resizebox{\columnwidth}{!}{
\centering
\begin{tabular}{lrccccc}
\toprule
  \multicolumn{1}{c}{\begin{tabular}[c]{@{}c@{}}Model\end{tabular}}                          & \multicolumn{1}{c}{\begin{tabular}[c]{@{}c@{}} Params\end{tabular}} & \begin{tabular}[c]{@{}c@{}}CNN\\ Unit\end{tabular} & \begin{tabular}[c]{@{}c@{}}RNN\\ Unit\end{tabular} & \begin{tabular}[c]{@{}c@{}}Hidden Unit\\per Pitch \end{tabular} & Note F1(\%) & \begin{tabular}[c]{@{}c@{}}Note F1\\ w/ Offset(\%)\end{tabular} \\ \hline
\multicolumn{1}{l|}{Base (PAR$_{compact}$)} & 2.7M & 48 & 48 & 48 & 95.98 & 84.27 \\
\multicolumn{1}{l|}{Middle}  & 1.5M & 32 & 32 & 32 & 95.50 & 82.32 \\  
\multicolumn{1}{l|}{Small}   & 653K & 16 & 16 & 16 & 90.83 & 69.08 \\  
\multicolumn{1}{l|}{Tiny}   & 299K & 8 & 8 & 8 & 81.31 & 16.47 \\  
\multicolumn{1}{l|}{Single bottleneck}  & 50K & 8 & 8 & 1 & 66.99& 15.98 \\  \midrule
\end{tabular}
}
\label{tab:PCmodel}
\end{table}

\subsection{Model Size Experiment}
Figure \ref{fig:model_param} shows a comparison of model sizes between the PAR model and the state of the art models. The PAR model have a similar number of parameters compared to other CRNN based models, while the PAR$_{compact}$ model has a reduced size of 86\% (from 19.7M to 2.7M) at the cost of 1.5\%p decrease in the onset F1 score and 2.9\%p decrease in the offset F1 score. We investigated even smaller models than the PAR$_{compact}$ model, progressively reducing the size of CNN units, RNN units, and hidden units per pitch. We found that model training tends to be very unstable when the unit sizes are 16 or less, but we were able to get results sometimes. Table \ref{tab:PCmodel} shows the performance trajectory. The models can maintain decent performance (e.g. above 90\% note onset accuracy) up to the ``small'' size, but they degrade dramatically in the ``tiny'' and ``single bottleneck' sizes.

\subsection{Model Latency Experiment}
We investigated the model performance by progressively reducing the model latency. The latency was controlled by reducing the filter widths in the Conv-FiLM blocks. Table \ref{tab:shortdelay} shows the performance of both the PAR and PAR$_{compact}$ models with different combinations of filter widths along the time dimension as a 6-tuple where left to right corresponds to bottom to top layers. Model parameters remained stable regardless of filter size. PAR model: 19.59M - 19.73M. Compact model: 2.54M - 2.68M. As expected, both note onset and offset accuracy decrease with  shorter time contexts. Interesting, except for the extreme case where all filter widths are set to 1 (128 ms latency), the models show small performance drops in the range of 160 ms to 320 ms latency. Between the two models, PAR$_{compact}$ has larger performance drops as the latency gets smaller. 


\begin{table}[!t]
\label{tab:shortdelay}
\caption{\textbf{Model Latency Experiment}}
\resizebox{\columnwidth}{!}{
\begin{tabular}{l|cc|cccccc}
\toprule
                                                                             & \multicolumn{1}{c}{}                                                            &                                                                           & \multicolumn{3}{c}{Note} & \multicolumn{3}{c}{\begin{tabular}[c]{@{}c@{}}Note w/ offsets\end{tabular}} \\ \hline
 & \multicolumn{1}{c}{\begin{tabular}[c]{@{}c@{}}CNN Filter\\ Widths\end{tabular}} & \multicolumn{1}{c|}{\begin{tabular}[c]{@{}c@{}}Latency\\ (ms)\end{tabular}} & R (\%)     & P (\%)    & F1 (\%)    & R (\%)                        & P (\%)                        & F1(\%)                        \\ \hline
\multicolumn{1}{l|}{\multirow{5}{*}{\rotatebox[origin=c]{90}{{\centering PAR}}}}                                                              & (1,1,1,1,1,1)                                                                   & 128                                                                       & 90.21     & 96.20    & 93.01     & 72.99                        & 77.67                        & 75.18                        \\
\multicolumn{1}{l|}{}                                                                            & (3,1,1,1,1,1)                                                                   & 160                                                                       & 94.47     & 98.81    & 96.55     & 83.52                        & 87.31                        & 85.34                        \\
\multicolumn{1}{l|}{}                                                                            & (3,1,3,1,1,1)                                                                   & 192                                                                       & 94.93     & 98.81    & 96.81     & 84.96                        & 88.40                        & 86.62                        \\
\multicolumn{1}{l|}{}                                                                            & (3,1,3,1,3,1)                                                                   & 224                                                                       & 95.14     & 98.71    & 96.87     & 85.69                        & 88.86                        & 87.22                        \\
\multicolumn{1}{l|}{}                                                                      & (3,3,3,3,3,3)                                                                   & 320                                                                       & 95.60     & 98.49    & 97.01     & 86.63                        & 89.21                        & 87.89                        \\ \hline

\multicolumn{1}{l|}{\multirow{5}{*}{\rotatebox[origin=c]{90}{{\centering PAR$_{compact}$}}}}                                                                            & (1,1,1,1,1,1)                                                                   & 128                                                                       & 76.31     & 96.76    & 84.25     & 41.18                        & 51.41                        & 45.18                        \\
\multicolumn{1}{l|}{}                                                                            & (3,1,1,1,1,1)                                                                   & 160                                                                       & 92.81     & 96.01    & 94.33     & 77.02                        & 79.60                        & 78.25                        \\
\multicolumn{1}{l|}{}                                                                            & (3,1,3,1,1,1)                                                                   & 192                                                                       & 93.44     & 98.31    & 95.77     & 79.65                        & 83.78                        & 81.63                        \\
\multicolumn{1}{l|}{}                                                                            & (3,1,3,1,3,1)                                                                   & 224                                                                       & 93.56     & 97.77    & 95.58     & 77.00                        & 80.43                        & 78.64                        \\
\multicolumn{1}{l|}{}                                                         & (3,3,3,3,3,3)                                                                   & 320                                                                       & 94.91     & 97.22    & 96.03     & 83.67                        & 85.69                        & 84.65                        \\ \hline
\end{tabular}
}
\end{table}


\begin{table}[!t]
\label{tab:crossdataset}
\caption{\textbf{Cross Dataset Evaluation.}}
\label{tab:CrossDataset}
\resizebox{\columnwidth}{!}{%
\begin{tabular}{@{}l|l|ccccc@{}}
\toprule
                           & Model         &\begin{tabular}[c]{@{}c@{}}MAESTRO\\V3\end{tabular} & MAPS & SMD  & \begin{tabular}[c]{@{}c@{}}Vienna\\Corpus\end{tabular}  & In-house\\ \hline
\multirow{4}{*}{\rotatebox[origin=c]{90}{{\centering Note}}}  & PAR               & \textbf{97.0}    & 79.9 & 95.2 & 93.5  &  96.1       \\
                           & PAR$_{compact}$     & 95.6    & 82.0 & 95.9 & 94.8  &  96.1    \\
                           & Regressing Onsets & 96.8    & \textbf{82.2} & \textbf{96.7} & \textbf{96.5}  &  \textbf{96.8}       \\
                           & Onsets and Frames (r) & 94.1    & 80.2 & 92.5 & 93.2  &  94.8       \\ \midrule
\multirow{4}{*}{\rotatebox[origin=c]{90}{{\centering Note /off}}} & PAR               & \textbf{88.0}    & 57.3 & \textbf{84.3} & 80.0  &  \textbf{88.8}       \\
                           & PAR$_{compact}$     & 85.1    & 60.3 & 82.9 & 76.5 &  88.0        \\
                           & Regressing Onsets & 84.7    & \textbf{60.9} & 83.3 & \textbf{83.4}   &  87.7     \\
                           & Onsets and Frames (r) & 79.8    & 55.1 & 76.2 & 67.2   &  83.4      \\ \bottomrule
\end{tabular}
}
\end{table}

\subsection{Cross Dataset Evaluation}
\label{sec:cross_dataset_eval}
We evaluated the proposed models (PAR and PAR$_{compact}$) on four piano music datasets other than MAESTRO. The unseen datasets during model training have different recording conditions and piano quality. Therefore, the results can show the generalization ability of the models. Table \ref{tab:crossdataset} shows the accuracy scores, comparing the proposed models with the Regressing Onsets model \cite{kong2021high} and the Onsets and Frames model \cite{hawthorne2018enabling}. Note that the two models compared were also trained in the same MAESTRO v3 training set and Onsets and Frames is the reproduced model. 
On Note F1 score, the Regressing Onsets model generally outperforms the proposed model and the Onsets and Frames model. This is likely because the regression loss proposed to mitigate the alignment errors in the audio-MIDI pairs of the MAESTRO dataset increases model generalization. Interestingly, the PAR$_{compact}$ model is the second best, achieving better performance than the PAR$_{compact}$ model. This seems to be due to a regularization effect by the small model size. On Note with offset F1 score, the trend is different. The PAR model become comparable to the Regressing Onsets model, archiving best performance on SMD and In-house. This is likely due to the enhanced recursive context designed to reduce note offset  missing. 

The three models generally show low performance on MAPS. We suspect that some of the errors may be due to annotation errors in the dataset itself, as reported in previous work \cite{gong2019analysis, hawthorne2017onsets}. Another possibility is that the acoustic difference between an upright piano and a grand piano is significant, given that the MAPS dataset is the only one played on an upright piano. 

\begin{figure*}[!htbp]
\centering
 \includegraphics[width=0.9\textwidth]{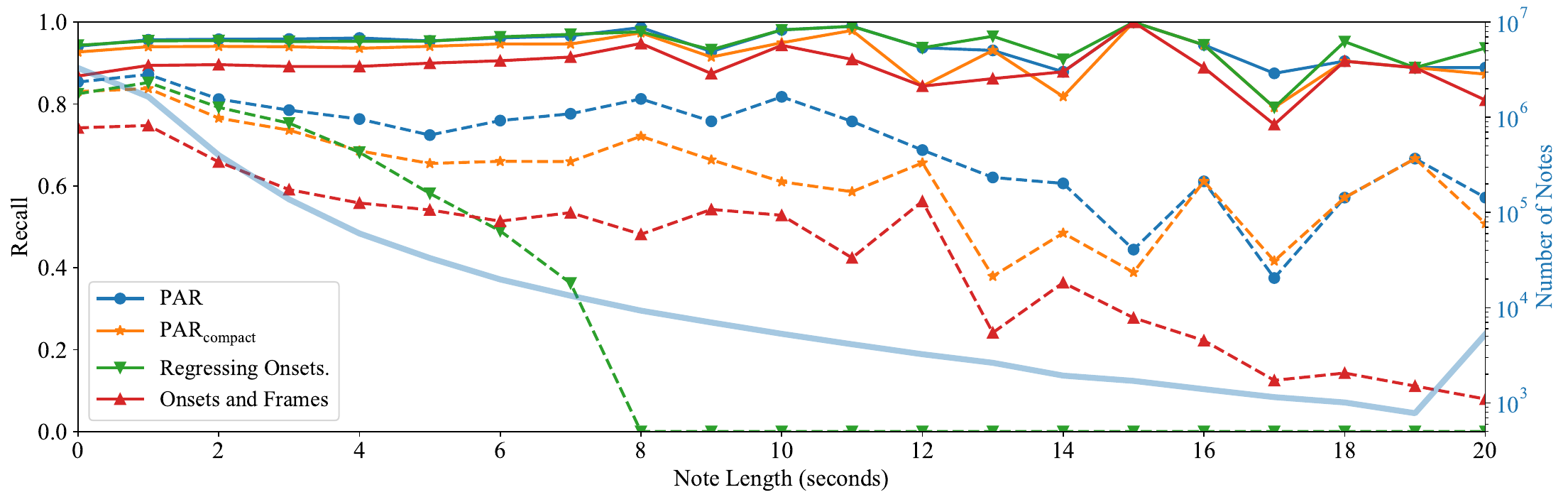}
 \caption{\textbf{Note Length Vs Note Recall.} Solid and dotted lines correspond to onset and offset, respectively. A thick light blue line indicates the number of notes in the training set (log scale).}
 \label{fig:lengtheval}
\end{figure*}


\subsection{Performance Analysis: Note Length View}
Note offset accuracy is likely to decrease for longer notes because there are more time frames to predict a note offset given a note onset. To address this issue, we incorporated the extended recursive context into the auto-regressive connection and demonstrated its effectiveness in the ablation study (section \ref{subsec:abalation}). We examine it further by analyzing the model's performance in the note duration view. Figure \ref{fig:lengtheval} plots the note with offset recall score over different note durations for the proposed models, the onsets and frames model, and the Regressing Onset model. While they have similar trends in note onset accuracy (solid lines), the two autoregressive models (PAR and PAR$_{compact}$) have a slower decline in the note offset recall than the two non-autoregressive models (dotted lines). In particular, the Regressing Onset model has a rapid decline in the recall performance for longer notes (dotted green line), being unable to estimating any note longer than 8s. This is presumably because the high time-resolution of the model (10ms per frame) hinders learning dependence between onset and offset in long notes. This result confirms that our proposed auto-regressive models are more reliable for longer notes.   


\label{subsec:pitch_gen}
\begin{figure*}[!htbp]
 \includegraphics[width=\textwidth]{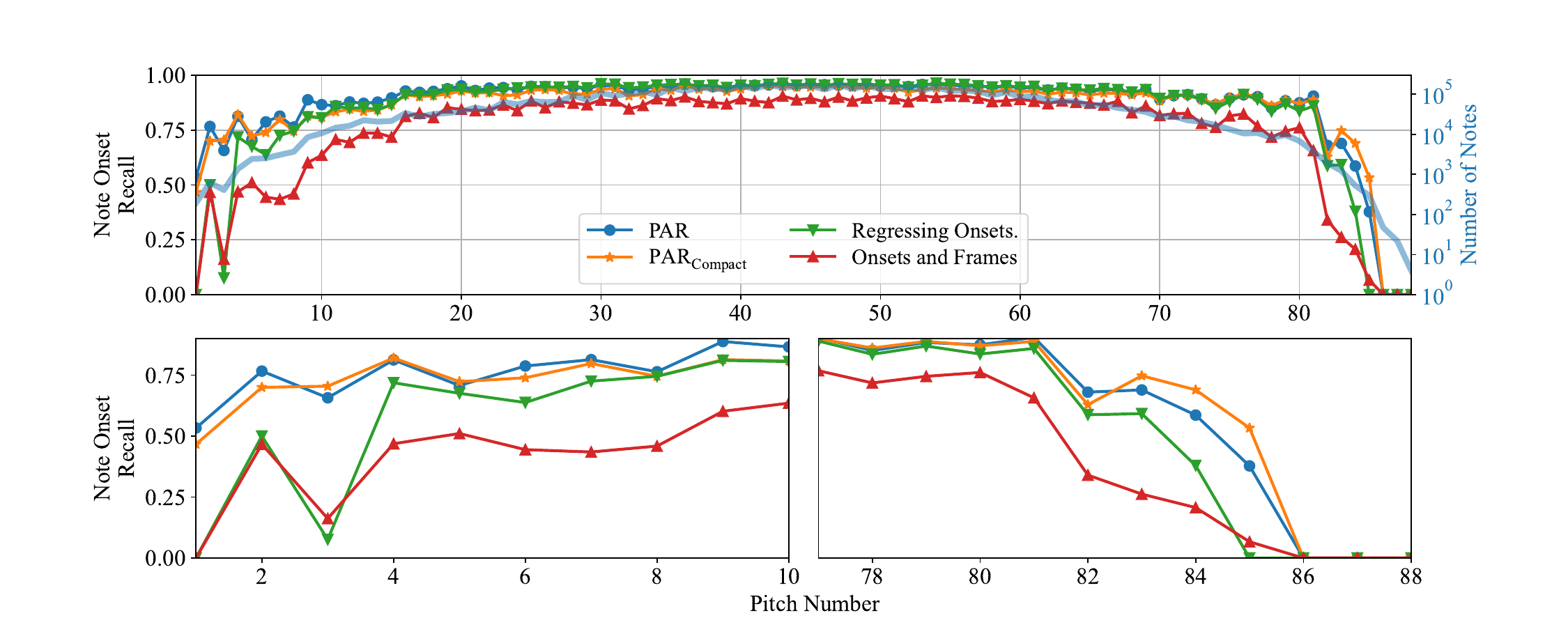}
 \caption{\textbf{Pitch Vs Recall.} The figure below is an enlarged view of both ends of pitches. The thick light blue line indicates the number of notes in the training set (log scale).}
 \label{fig:pitcheval}
\end{figure*}

\subsection{Performance Analysis: Pitch Range View}
The distribution of notes played in most piano music, including MAESTRO, is unbalanced; low or high notes are played relatively less frequently than middle notes. The imbalance distribution affects the transcription performance; the note accuracy is generally proportional to the note frequency. We investigate the effect of components in our proposed models. Figure \ref{fig:pitcheval} plots note onset recall against note number, comparing the four models. As expected, note onset recall decreases at both ends of the pitch range. However, the PAR and PAR$_{compact}$ models have much higher recall at both ends. This is more easily recognizable in the magnified plots of the very low and high pitch range in the bottom of Figure \ref{fig:pitcheval}. We particularly attribute this to the pitchwise LSTM because the parameters of the pitchwise LSTMs are shared across all pitches and therefore low and high pitch notes harness the note state information from  middle pitch notes. 

\begin{figure}[!tp]
 \includegraphics[width=\columnwidth]{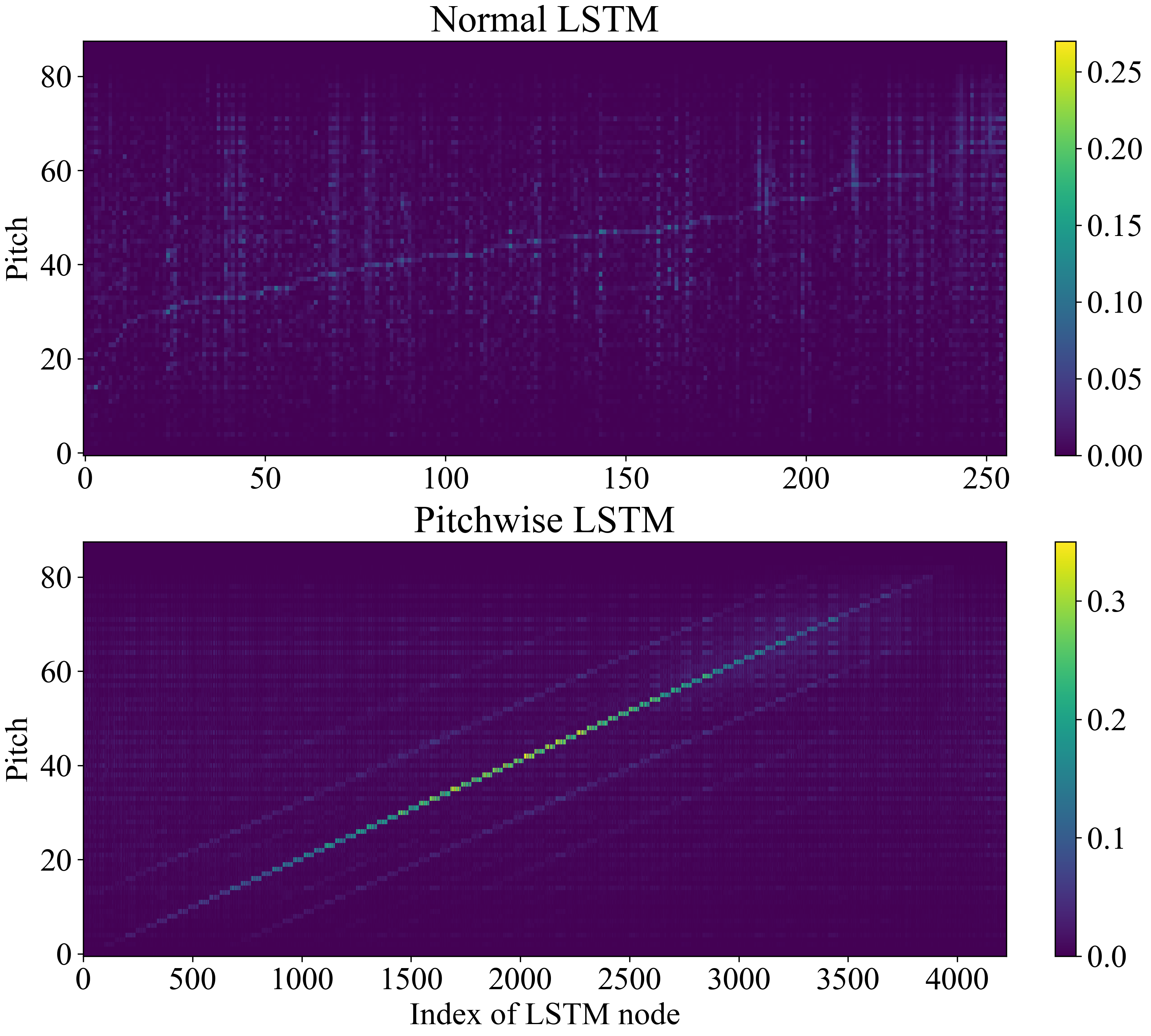}
 \caption{Mutual information between the LSTM activations (second layer) and the notes in each pitch. PAR with normal LSTM model ((b) in Tab. \ref{tab:PAR_ablation} and PAR model with pitchwise LSTM (f).}
 \label{fig:mi}
\end{figure}

We further validate this ``pitch generalization'' effect by visualizing the mutual information between LSTM neurons and pitches. Figure \ref{fig:mi} contrasts the mutual information in the normal LSTM (model (c) in Table \ref{tab:PAR_ablation}) and the second LSTM layer of the pitchwise LSTM. To compute the mutual information, we randomly sampled 20,000 frames from the training dataset and used a function in the scikit-learn\footnote{\url{https://scikit-learn.org/stable/modules/generated/sklearn.feature_selection.mutual_info_classif.html}} \cite{scikit-learn} package based on entropy estimation from k-nearest neighbor distances between continuous variables (LSTM layer activations) and discrete variables (pitch labels) \cite{ross2014mutual}. The visualization shows that a significant proportion of the neurons in the normal LSTM are assigned to a single pitch. The number of nodes assigned to each pitch was roughly proportional to the number of notes in that pitch in the dataset. As a result, many pitches on both ends were not assigned to any neurons in the normal LSTM. In contrast, our PAR model had an equal and large number of neurons assigned to each pitch, regardless of the number of notes in each pitch, resulting in better generalization across pitches.

\section{Conclusions}
We presented novel CRNN models for piano transcription that can simultaneously achieve high-performance, lightweight, and real-time inference. Through the comprehensive experiments, we show that the the  proposed models are comparable to previous state-of-the-art models in performance, while having a compact model size. We validated the effectiveness of the proposed components (frequency-conditioned FiLM, pitchwise LSTM, and enhanced recursive context) through an ablation study. Also, we analyzed the results in the view of note length and pitch range to elucidate the effect of the proposed components further.   



One of the limitations in this study is that the evaluation on unseen piano dataset (Section \ref{sec:cross_dataset_eval}) was conducted only when the models are trained with a single training set (MAESTRO V3). In order to validate the generalization power and the effectiveness on real-world examples, we need to augment the dataset \cite{hawthorne2018enabling} or leverage unlabeled datasets with the aid of semi-supervised \cite{cheuk2021reconvat} or unsupervised learning method \cite{choi2019deep, gfeller2020spice}. In addition, our models do not include an inter-pitch language model. While our intra-pitch-focused models achieved high performance, inter-pitch connections are still useful if the time resolution can be reduced \cite{Ycart2019BLENDINGAA}. Finally, we look forward to the development and implementation of a system capable of analyzing a performance or providing real-time analysis in real-world recording scenarios.







%

\section*{Acknowledgment}
This work was supported by the National Research Foundation of Korea (NRF) grant funded by the Korea government (MSIT) (No. NRF-2023R1A2C3007605).

\ifCLASSOPTIONcaptionsoff
  \newpage
\fi



\bibliographystyle{IEEEtran}
\bibliography{jrnl}
%

%

\begin{IEEEbiography}
[{\includegraphics[width=1in,height=1.25in,clip,keepaspectratio]{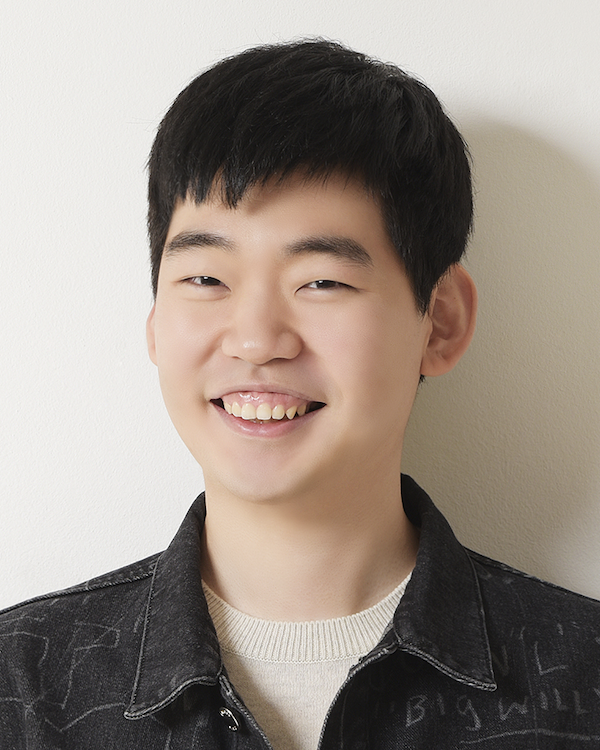}}]{Taegyun  Kwon} is currently working as an Post-doc in the Graduate School of Culture Technology at the Korea Advanced Institute of Science and Technology (KAIST) in South Korea. He obtained his Ph.D. and M.S. degrees in culture technology, and B.S. in Physics from Korea Advanced Institute of Science and Technology (KAIST). His research primarily focuses on music transcription with deep learning approach. His research interests also include expressive performance modeling and human computer interaction in musical performance.
\end{IEEEbiography}

\begin{IEEEbiography}[{\includegraphics[width=1in,height=1.25in,clip,keepaspectratio]{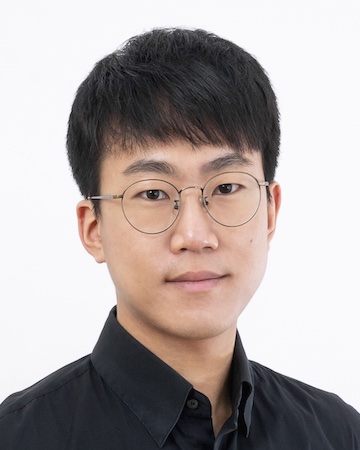}}]{Dasaem Jeong} is currently working as an Assistant Professor in the Department of Art \& Technology at Sogang University in South Korea since 2021. Before joining Sogang University, he worked as a research scientist in T-Brain X, SK Telecom from 2020 to 2021. He obtained his Ph.D. and M.S. degrees in culture technology, and B.S. in mechanical engineering from Korea Advanced Institute of Science and Technology (KAIST). His research primarily focuses on a diverse range of music information retrieval tasks, including expressive performance modeling, symbolic music generation, and cross-modal generation.
\end{IEEEbiography}


\begin{IEEEbiography}
[{\includegraphics[width=1in,height=1.25in,clip,keepaspectratio]
{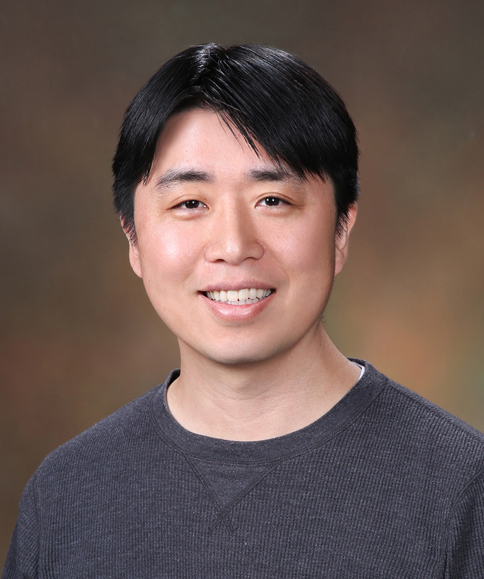}}]{Juhan Nam} is an Associate Professor in the Graduate School of Culture Technology at the Korea Advanced Institute of Science and Technology (KAIST) in South Korea. Prior to joining KAIST, he worked as a Staff Research Engineer at Qualcomm from 2012 to 2014 and as a Software/DSP Engineer at Young Chang (Kurzweil) from 2001 to 2006. He received his Ph.D. in Music from Stanford University, where he studied at the Center for Computer Research in Music and Acoustics (CCRMA). He also holds an M.S. degree in Electrical Engineering from Stanford University and a B.S. degree in Electrical Engineering from Seoul National University. His research interests include the application of digital signal processing and machine learning to music and audio. He is a member of the IEEE.
\end{IEEEbiography}




\end{document}